# Open Data Sharing in Clinical Research and Participants Privacy: Challenges and Opportunities in the Era of Artificial Intelligence

## Authors
Shahin Hallaj[1,2], Anna Heinke[1,2], Fritz Gerald P. Kalaw[1,2], Nayoon Gim[3,4,5], Marian Blazes[3,4], Julia Owen[3,4], Eamon Dysinger[6], Erik S. Benton[7], Benjamin A. Cordier[8], Nicholas G. Evans[9], Jennifer Li-Pook-Than[10], Michael P. Snyder[10], Camille Nebeker[11], Linda M. Zangwill[1], Sally L. Baxter[1,2], Shannon McWeeney[8], Cecilia S. Lee[3,4], Aaron Y. Lee[3,4], Bhavesh Patel[12,*], on behalf of the AI-READI Consortium

## Affiliations

[1]Division of Ophthalmology Informatics and Data Science, Hamilton Glaucoma Center, Viterbi Family Department of Ophthalmology and Shiley Eye Institute, University of California, San Diego, La Jolla, California
[2]Division of Biomedical Informatics, Department of Medicine, University of California, San Diego, La Jolla, California
[3]Department of Ophthalmology, University of Washington, Seattle, Washington
[4]Roger and Angie Karalis Johnson Retina Center, Seattle, Washington
[5]Department of Bioengineering, University of Washington, Seattle, Washington
[6]Oregon Health & Science University
[7]Oregon Clinical and Translational Research Institute, Oregon Health & Science University
[8]Knight Cancer Institute, Oregon Health & Science University
[9]Department of Political Science, University of Massachusetts Lowell
[10]Genetics Department, Stanford University, Stanford, California
[11]Herbert Wertheim School of Public Health and Human Longevity Science, University of California, San Diego, La Jolla, California
[12]FAIR Data Innovations Hub, California Medical Innovations Institute, San Diego, California

*Corresponding author
Bhavesh Patel
Research Professor
FAIR Data Innovations Hub
California Medical Innovations Institute
11107 Roselle St.
San Diego, CA 92121
T: (510) 604-2815, Email: bpatel@calmi2.org

**Unstructured Summary (149/150 words)**

Sharing clinical research data is key for increasing the pace of medical discoveries that improve human health. However, concern about study participants' privacy, confidentiality, and safety is a major factor that deters researchers from openly sharing clinical data even after deidentification. This concern is further enhanced by the evolution of artificial intelligence (AI) approaches that pose an ever-increasing threat to the reidentification of study participants. Here, we discuss the challenges AI approaches create that are blurring the lines between identifiable, and non-identifiable data. We present a concept of pseudo-reidentification, and discuss how these challenges provide opportunities for rethinking open data sharing practices in clinical research. We highlight the novel open data sharing approach we have established as part of the AI-READI (Artificial Intelligence Ready, and Exploratory Atlas for Diabetes Insights) project, one of the four Data Generation Projects funded by the National Institutes of Health Common Fund's Bridge2AI Program.

## Introduction

Sharing scientific research data is a cornerstone for advancing science, and accelerating discoveries. Clinical research is a prime illustration where broad data sharing can rapidly drive innovations that benefit patient care. For instance, the widespread sharing of genomic data has enabled rapid advancement in cancer research.[1] During the COVID-19 pandemic, rapid data sharing enabled a quick understanding of the virus.[2] Clinical data sharing efforts have particularly grown over the past decade, driven by the rise of initiatives such as the Findable, Accessible, Interoperable, Reusable (FAIR) principles.[3]

Despite these efforts, persistent challenges hinder effective data sharing, and reuse. Researchers, and institutions hesitate to make data openly available to external investigators. Concerns about participant privacy, data security, and potential misuse limit the sharing of valuable datasets.[4] Study participants also have similar concerns. In a 2024 nationwide online survey of adults in the United States, a majority of respondents felt relatively at ease sharing data with healthcare providers.[5] The study also highlighted a critical tradeoff between the push for open science to improve clinical outcomes, and public health, and the need to honor patient privacy, autonomy, and trust in data collection, and use. Enhancing the transparency, and governance of data sharing practices is necessary to maintain participant confidence, and willingness to contribute to research.[5] Moreover, the advent of increasingly sophisticated artificial intelligence (AI) tools exacerbates the reidentification risk, blurring the line between identifiable, and non-identifiable data types, and raising questions about participant privacy.[6]

In this paper, we examine the evolving challenges, focusing on the privacy risks introduced by AI. We discuss examples of data types currently not deemed identifiable, in which, through AI-driven analysis, uniqueness can be inferred without actual reidentification, a process we term "pseudo-reidentification". Finally, we introduce a novel data sharing approach developed within the AI-READI (Artificial Intelligence Ready, and Exploratory Atlas for Diabetes Insights) project, part of the National Institutes of Health (NIH) Common Fund's Bridge2AI Program,[7,8] aiming to safeguard participant privacy while preserving the spirit of openness, and collaboration that propels clinical research forward.

## Regulatory Landscapes for Data Protection, and Privacy

A summary of concepts related to data privacy is provided in a glossary in Table 1.

### HIPAA, PII, and PHI

Personal identifiable information (PII) refers to any information that links to an individual.[9] A protected health information (PHI) is collected for the provision of healthcare services, and protected by the Health Insurance Portability, and Accountability Act (HIPAA) of 1996.[10] The term "deidentification" originates from the HIPAA, which involves the identification, and removal of PHI from data. According to the HIPAA Privacy Rule, if health information is deidentified, it is not considered PHI, and deidentified datasets may be shared more easily. While HIPAA does not regulate deidentified data, other ethical, legal, or institutional constraints may still apply. A dataset can be deidentified under HIPAA following one of two methods: *Safe Harbor,* or *Expert*

*Determination*. The Safe Harbor method requires 18 health information elements (listed in Table 2) to be removed from the dataset.[11] In Expert determination method, an expert certifies that the risk of reidentification is minimal, regardless of the specific method used.[12] In some cases, however, the use, and disclosure of PHI are needed for research, public health, or healthcare operations. In such cases, datasets with PHI can be shared as "limited datasets" with proper restrictions, and security measures, including a data use agreement between the data provider, and the data user.[13] HIPAA was designed as flexible guidance rather than strict regulation, enabling it to adapt over time as technology evolves. As a result, the determination of what constitutes PHI (Table 2) can vary among institutions, and organizations.

**General Data Protection Regulation (GDPR)**
In the European Union (EU), personal data are regulated by the General Data Protection Regulation (GDPR), which came into effect in 2018.[14] The GDPR is widely regarded as one of the strictest data protection frameworks globally, and is considered highly "data subject-centric," protecting the privacy rights of all individuals in the EU-not just patients. The GDPR applies to any organization, whether inside, or outside the EU, that offers goods, or services to, or monitors the behavior of, individuals located in the EU.[15] The regulation requires data controllers, and processors to implement robust safeguards to ensure privacy, such as data pseudonymization, or encryption.[15] It distinguishes pseudonymization from anonymization (Table 1).[16] Pseudonymized data is still considered personal data under the GDPR, and remains subject to its requirements. In contrast, once data are truly anonymized, they are no longer regulated by the GDPR. Under the GDPR, any information that can directly, or indirectly identify an individual, including biometric data, is classified as personal data. Processing personal data is only permitted if there is a lawful basis, such as explicit consent, performance of a contract, compliance with a legal obligation, protection of vital interests, the performance of a task carried out in the public interest, or legitimate interests pursued by the controller, or a third party. The consent must be freely given, specific, informed, and unambiguous, and individuals must be able to withdraw consent at any time. In addition, participants have the right to data portability, allowing individuals to receive their data in a commonly used, machine-readable format, and to transmit that data to another data controller.[17] When third parties process data on behalf of a controller, data processing agreements must be established to ensure compliance.[18] The approach to de-identification, and secondary use of data may vary across jurisdictions, but the GDPR sets a high standard for privacy, and data security.

## Pseudo-reidentification in the Era of AI

**Reidentification vs. Pseudo-reidentification**
Traditional reidentification involves linking deidentified data back to a known individual by leveraging external identifiers, or datasets. In studies using deidentified open-source databases, only the primary investigators (PI) who collected the data may possess access to the identifiers linking these records to PHI. Quasi-identifiers (QI) may also put research participants at risk. QIs are data elements that do not directly identify participants but can be used for reidentification when linked to other sources of information.[19] Traditional examples of QIs include: birth weight, behavioral data, sex, profession, total income, minority status, locations, spoken languages, ethnicity, education, marital status, criminal history, disability, dates, codes (e.g., diagnosis, procedure, or adverse event codes), and birth plurality.[19] While HIPAA's Safe Harbor method removes many identifiers, and some QIs, some QIs may remain in deidentified datasets. Linking QIs to external information that contains PHI may also lead to reidentification.[19] Despite advancements in health cybersecurity, and infrastructure, the risk of malicious attacks exposing these identifiers remains a concern.[20] While such breaches could expose sensitive patient data, the overall risk remains lower than breaches involving data stored by HIPAA-covered entities (e.g., insurance companies, healthcare systems, or EHR providers). Moreover, the likelihood of reidentification by third parties (i.e., other than covered entities, or project PIs) is minimal.[21]

However, AI can discern unique patterns in data elements that, while not explicitly tied to an individual, still imply uniqueness (Table 3). For example, AI models may detect unique structural, physiological, or behavioral patterns that, while not directly tied to a name, or ID, are unique enough to single out an individual in a dataset.[22] We introduce the term 'pseudo-reidentification' to describe this identification of unique data patterns that do not directly link to an individual, but could be linked if identifiers become available (Figure 1). The only step between pseudo-reidentification, and reidentification is the linkage of the pseudo-reidentified data elements to external identifiers (e.g., PHI, or PII). While this may be unlikely if only PIs have access to identifiers, pseudo-reidentification remains a concern as technological advances may make this linkage more feasible. These evolving risks challenge current regulatory frameworks, which may not fully account for pseudo-reidentification. Below, we review studies assessing pseudo-reidentification for commonly collected data types in clinical research, and traditionally not considered PHI (Table 3).

## Practical Insights into Pseudo-Reidentification by Modality

### *Wearable Fitness-Tracking*
Fitness-tracking devices record signals such as heart rate (including variability, and pattern), step count, gait (including fixed time durations, step cycles, and walk cycles), metabolic equivalent of task, energy expenditure, and exercise parameters recorded via the global positioning system (GPS), and accelerometer.[23] These wearables use biometric sensors to continuously monitor physiological signals, leveraging each individual's unique baseline to enable accurate authentication and early detection of health changes or unusual health activity. Therefore, it can perform real-time detection of signals in a non-invasive way, making the data acquisition convenient. However, considering the unique individual activity habits and constant monitoring of data elements including photoplethysmography (which detects blood volume changes in the microvascular bed of tissue), heart sounds, movement patterns, and heart rate raise concerns about the reidentifiability of this data type.[24] Researchers applied various machine learning methods, such as support vector machines (SVMs), and random forests (RFs), neural networks, and deep learning (DL) strategies.[25] . All studies reported high accuracy from deidentified wearables information, noting that pseudo-reidentification is possible with small data fragments.[26]

*Continuous Glucose Monitoring*
Continuous glucose monitors (CGMs) track blood glucose levels, enabling improved monitoring and informed diabetes management.[27] CGMs allow prompt detection of glycemic changes during acute stress by notifying the user, or healthcare provider, ensuring proactive management of such events.[28] CGMs generate a substantial amount of data, which is synchronized, stored, and shared across different platforms. Deidentified CGM data from multiple study groups are accessible for secondary use,[29] and are not considered PHI, despite significant cybersecurity implications.[30] Some manufacturers may gather PIIs, such as the user's internet protocol (IP) address, network accessibility, internet service, browser, and their activities, which can eventually lead to reidentification.[28] Although manufacturers claim to deidentify CGM data, how they perform deidentification is not usually mentioned. This raises privacy, and security concerns for CGM users. CGM data can be used to pseudo-reidentify individuals using ML algorithms (Table 3).[30] Reported accuracy may reach as high as 86% in pseudo-reidentifying CGM users. With the growing number of patients with diabetes wearing CGMs, as well as CGM manufacturers, privacy concerns of consumers are increasing. Recognizing this risk, the Institute of Electrical, and Electronics Engineers Standards Association published standards to help stakeholders develop more secure wireless diabetes devices.[31]

*Electrocardiogram*
Electrocardiogram (ECG) data is a record of the electrical signals generated by cardiac rhythm, and activity. ECG data is not considered PHI, and several deidentified datasets are publicly available online, with the rationale that their linkage to PII, or PHI is unlikely.[32] However, studies reporting on biometric recognition using ECG date back to 2001,[33] when Biel et al. applied soft independent modeling by class analogy (SIMCA) on features extracted from 12-lead ECG records to link subsequent ECGs taken in the same visit with their baseline ECG. Nowadays, off-the-person devices (i.e., wearable devices)[34] may also be used for this purpose, in addition to the classic on-the-skin (i.e., on-the-person) 12-lead ECG. The accuracy can range between 75-100%,[35] depending on the used device, test duration, and test intervals. Similar to advancements in databases, and hardware, analytical approaches have evolved. Earlier approaches included manual feature extraction,[33] and principal component analysis. Recently, more studies report AI applications,[36] including DL-based pseudo-reidentification via uni-, or multimodal modeling[37] of ECG along with biometrics such as fingerprints. All studies have used publicly available databases, and PHI is not available in any of the mentioned databases.

*Retinal Imaging*
Retinal imaging involves capturing images of the retina, the light-sensitive tissue at the back of the eye, using advanced technologies such as optical coherence tomography, color fundus photography, and OCT angiography. These techniques provide high-resolution, cross-sectional, or wide-field views, aiding the diagnosis, and management of various retinal conditions. Several publicly accessible retinal imaging datasets are available online.[38] The distinct vascular patterns in retinal scans may act as a potential identifier for individuals, posing privacy concerns. The published body of the literature suggests that retinal images are pseudo-reidentifiable (Table 3). The accuracy of pseudoreidentification using retinal images is similar to the previously discussed modalities, ranging from 95 to 99%, suggesting that uniqueness alone can be detected, but can not be linked to PHI without external identifiers. The American Academy of Ophthalmology advises against the consideration of retinal images as biometric identifiers for clinical research.[39] Unlike established biometric identifiers, such as fingerprints, or iris scans, retinal imaging quality varies significantly due to differences in equipment, technique, and patient conditions. Moreover, the features detected in retinal images are not static; they can change over time due to aging, disease progression, or treatment, further complicating their reliability as a stable individual identifier.[40]

**Hidden Pathways to Reidentification: Navigating Modern Data Misuse**
Although the likelihood of identifying an individual solely from the data types described above may seem relatively low, it is not negligible. The mentioned data elements are not traditionally known as identifiers or QIs. However, their linkage to external sources of information containing PHI may lead to actual reidentification. Advances in AI-driven analytics mean that even fragments of non-traditional biometric data, such as aggregated wearable metrics, ECG signals, or retinal patterns, could be leveraged for malicious purposes.[41] Ill-intended individuals may devote significant time, and resources to parsing the web, obtaining additional context from social media, online medical forums, or leaked datasets, and combining these disparate elements to infer a person's identity, sensitive data, or health status.[42] These priorities are not irreconcilable, but require tiered, auditable, and governed access protocols with explicit awareness that each reuse incrementally draws down a finite privacy reserve.[43] Further, digital health companies are dedicating increasing resources to acquire real-world data from different populations. Data leak, misconduct, or reidentification attempts by companies that already have access to PII can be another form of data misuse.[44] Moreover, as data-sharing practices expand, the inevitable presence of data brokers, dishonest third parties, or data enthusiasts increases the risk that reassembled fragments of deidentified information could be used to discriminate, stigmatize, or exploit individuals.[45] In other words, the risks, though not prominent, are real enough to demand thoughtful data protection, governance, and oversight.[18]

## Limitations of Current Data Sharing Methods
In light of these new risks, it is necessary to reexamine existing data sharing methods. These methods can be broadly classified into three categories: 1) Open sharing relying on deidentification, 2) Controlled access, and 3) Enclave-based access.[38]

Open data sharing presents the simplest way to maximize data accessibility. Researchers may, or may not be required to register, and share their information to access the data.[46] Data is shared under a data reuse license that is typically permissive, such as the Creative Commons Attribution 4·0 International (CC-BY 4·0), which allows reuse for any purpose. Data access is often as easy as clicking a "Download" button. Examples include autonomic nervous system-related datasets available on the NIH SPARC program's repository, and neuroimaging datasets available on the OpenNeuro repository.[47,48] They rely on the researchers sharing the data to make sure it does not contain any PHI, through deidentification. These methods may have limited protection against data misuse, as there is little legal framework for tracking, and reinforcement compared to more controlled methods. Depending on the strength of the legal framework, data misuse in this scenario may still have reputational, or legal consequences.

Controlled access methods typically require submitting a data access application, which is reviewed by a committee. Controlled access includes centralized, and decentralized (or federated) approaches. Centralized approaches usually include management of data from multiple sources in a single, centralized repository, and then granting access to the users. In contrast, decentralized approaches distribute data management across multiple sources, each handling its requests, and agreements. This can increase agility, enable more local control, and may foster higher data sharing rates, and faster research outputs, but may require more resources, and coordination, and can introduce inconsistencies in access procedures.[49] If access is requested for PHI elements, a Data Usage Agreement (DUA) is established between the sharing, and receiving entities. Accessors may be required to pay for registration, and data access to support the sustainability of the data sharing approach.[46] Examples of controlled access methods include datasets from the dbGaP repository, and the UK Biobank.[50,51] Controlled access may be viewed as a necessary compromise to protect participants' privacy, still enabling data reuse.[52] While reducing the risk of data misuse through vetting data accessors, these methods may go against the spirit of open science as they could exclude certain individuals from accessing the data (e.g., those not affiliated with a trusted institution). It can also delay data access, and present sustainability risks if the data access committee is unable, or unwilling to continue its duties.

Enclave-based access methods consist of granting access to the data within a secure storage called an enclave, where the data accessors must perform their analysis without the ability to take the data out. Getting access to the enclave usually also involves submitting a data access application. Examples of enclave access methods include data from the *All of Us* Research Program Researcher Workbench, and the N3C COVID Enclave Data.[38,53] This may be the most secure approach for protecting participant privacy, and preventing data misuse. However, the drawbacks of this approach may be the exclusion of certain individuals from accessing the data, especially those with limited knowledge of working in enclaves. In addition, this approach could be cost-prohibitive as it may require providing computational resources to users. It can also limit the ability to combine, and analyze data from different studies.

Overall, data openness reduces moving from the open sharing to controlled access, and enclave-based categories, while participants' security, and the cost associated with long-term access to the data correspondingly increase. There is a necessity for a method that protects the participants' privacy, especially against the increasing threats caused by AI, without compromising data openness. Recognizing that current frameworks fall short, and can foster false-security, the AI-READI project has introduced a novel open data sharing approach. This approach is designed to preserve data usability for research while establishing more robust safeguards against misuse, and unauthorized reidentification attempts.

## The AI-READI Open Data Sharing Method

AI-READI is one of the four Data Generation Projects funded by Bridge2AI, an NIH Common Fund Program aimed at setting the stage for widespread adoption of AI in health research. The project seeks to create a flagship dataset to provide critical insights into Type 2 Diabetes Mellitus (T2DM), including salutogenic pathways to return to health.[7,8] Data is collected from individuals with, and without T2DM, and harmonized across three data collection sites in the United States. The composition of the dataset consists of a multimodal array of data, including survey data, physical measurements, cognitive testing, vision testing, laboratory values, retinal imaging, ECG data, continuous glucose monitors, physical activity monitors, and home environmental sensors.[8] Participant enrollment for data collection started in the summer of 2023. A total of 4,000 participants are planned to be enrolled by the end of the project in November 2026. A major goal is to broadly share this multimodal dataset such that it is ready for AI/ML-related applications. The second version of the dataset, containing data from 1067 participants, was shared in November 2024.[54]

AI-READI participants provide informed consent emphasizing data sharing practices, and privacy protections. In research, the informed consent process is designed to facilitate understanding of what data will be collected and how data will be shared and used. It explicitly addresses potential risks, including data privacy concerns, and, outlines the measures taken to deidentify data, and maintain confidentiality. Participants are also made aware that while they can withdraw from the study at any time, data that has already been shared may remain in public, or controlled-access databases. This dataset has two sets. The first set is a public set that can only be used for T2DM-related research, and excludes the following data elements: ZIP code, genetic sequence, health records, motor vehicle accident reports, medications, sex, race/ethnicity. The decision to allow use of the public set only for T2DM-related research is meant to align with the consent. The public set is free from PHI, and could be shared under a method from the open access category described previously. Additionally, withholding race/ethnicity, or sex from the public set is intended to prevent findings that may stigmatize certain groups. The second set is a controlled set containing all the collected data, and can be used for any approved purpose. The controlled set requires a data usage agreement (DUA) for access. We took this opportunity to design a novel data access approach, considering pseudo-reidentification risks.

This novel data access process is implemented in FAIRhub, a novel data sharing platform designed to maintain open access while ensuring participants' privacy. We designed this model using a Swiss-cheese approach of open data sharing, and it contains several layers that may not be foolproof to protect participant privacy on their own, but can significantly reduce such risk when put in sequence (Figure 2). The first layer consists of authenticating with an identity-verified system, which enables getting information about the person accessing the data (name, institutional email, affiliation) in a reliable way. The user is informed that their name, email, and intended use of the dataset are saved in the FAIRhub database, and are visible to the public on the project website. Currently, the authentication process for data access on FAIRhub is done through CILogon, which is an open-source identity, and access management platform operated by the National Center for Supercomputing Applications at the University of Illinois. Users from many institutions across the globe can authenticate using this platform. Its major limitation is the inability to provide attestation for non-academic individuals because it federates with known identity management providers. We are exploring alternative verified identity providers to ensure secure access for researchers with appropriate credentials, and to promote responsible use of the dataset by the end users. The second layer consists of reading the license terms, and agreeing to adhere to them. Identifying a gap in commonly used data sharing licenses such as CC-BY-4·0, we have established a new license that allows reuse of data for research, or commercial purposes but includes certain restrictions in place to protect the privacy of study participants.[55] Additionally, the license explicitly prohibits data resharing (excluding with collaborators at the same institution), using models that remember the dataset, and attempting to reidentify the participants in any way. The third layer consists of attesting word-by-word to the major requirements mentioned in the license. This is intended to reinforce the requirements, and create a social contract that targets the individual user (while the License targets institutions). The fourth layer consists of describing the intended use of the data. This description is publicly posted on the dataset's landing page on FAIRhub along with the user's full name to provide full transparency about the use of the dataset, especially to the study participants, who can see what their data is used for. The fifth layer consists of watermarking the data. Watermarking is currently performed at the user level, meaning unique, and traceable watermarks on each file are associated with each user accessing the data (based on their identity obtained through layer 1). This enables tracing of any source of data leaks in the future, and allows individual attribution of any misuse, ensuring that anyone attempting to use the data outside of what the license permits will be held accountable. When ready, the user receives an email with the download instructions at their email address associated with the verified ID system they used to log in to FAIRhub, adding yet another layer of security.

Overall, we have designed a multi-step process that integrates several layers of protection from misuse but remains accessible, rapid, and autonomous, thus maintaining data openness while enhancing the privacy protection of the participants from current, and future threats posed by evolving AI approaches. Our requirement for identity verification, attestation, and public disclosure of user information could still deter some users, or introduce barriers compared to truly open access.

## Discussion

Data sharing is critical for advancing science. However, it may introduce inherent risks to study participants. The line between protected data, and data with the potential for reidentification is not always clear, becoming increasingly blurred with the advent of powerful AI approaches. Therefore, different standards, rules, and policies are implemented at various levels, including research groups, institutional, state, national, and sometimes even continental levels. For example, some data elements may be considered high-risk yet shareable under certain conditions at one institution, while another institution might classify the same data as low-risk, and allow open sharing.

We reviewed the most widely used data-sharing regulations, highlighted their limitations, and explored the concept of pseudo-reidentification. Pseudo-reidentification is possible through ECG, CGM, wearable, or retinal images using advanced AI techniques. While still a step away from reidentification, this shows how evolving technologies represent an ever-increasing risk to participants' privacy. We postulate that new data sharing approaches are required to mitigate these risks. Accordingly, we presented the new approach we have implemented in the AI-READI project, using a Swiss-cheese model for open data sharing. We provided the rationale behind our approach, aiming to reduce risks to participants' privacy while maintaining data openness.

The Swiss-cheese method of open data sharing is not intended to be a fixed method but to evolve with the addition, or removal of layers, to keep up with evolving privacy risks. As the current implementation is being tested by users accessing the AI-READI dataset through FAIRhub (422 dataset access as of April 2025), we will aim to identify the limitations of the current layers, and address them by exploring advanced technologies, such as blockchain-based audit logs, which offer transparent data tracking. We will also investigate under which circumstances participants are willing to incur different levels of pseudo-reidentification risk, and for which use cases.

We aim for this method to strengthen trust among researchers and study participants in openly shared datasets. We also hope our method will be adopted by other projects, either in its current form or as a foundation for developing new approaches that enhance participant protection while preserving data openness.

## Search strategy and selection criteria

References for this Review were identified through searches of PubMed, arXiv, and IEEE (Institute of Electrical, and Electronics Engineers) archive with the search terms "data sharing", "reidentification", "pseudo-reidentification", "research participant", "privacy", "biometric", "retinal imaging", "ECG", "wearable fitness tracking", and "continious glucose monitoring" from 2000 until April, 2025. Articles were also identified through searches of the authors' files. Only papers published in English were reviewed. The final reference list was generated based on originality, and relevance to the broad scope of this review.

## Acknowledgments
This work was supported by the NIH through grants OT2OD032644, and T32EY026590. We thank the Microsoft AI for Good Lab for supporting the cloud services needed for the project.

## Competing interests
The authors declare no competing interests.

## References

1   Clinical Cancer Genome Task Team of the Global Alliance for Genomics and Health, Lawler M, Haussler D, *et al.* Sharing clinical and genomic data on cancer - the need for global solutions. *N Engl J Med* 2017; **376**: 2006–9.

2   Moorthy V, Henao Restrepo AM, Preziosi M-P, Swaminathan S. Data sharing for novel

coronavirus (COVID-19). *Bull World Health Organ* 2020; **98**: 150.

3   Wilkinson MD, Dumontier M, Aalbersberg IJJ, *et al.* The FAIR Guiding Principles for scientific data management and stewardship. *Sci Data* 2016; **3**: 160018.

4   Data Sharing Concerns. Rethinking Clinical Trials. 2017; published online June 26. https://rethinkingclinicaltrials.org/chapters/dissemination/data-share-top/data-sharing-concerns/ (accessed April 9, 2025).

5   Niño de Rivera S, Masterson Creber R, Zhao Y, *et al.* Public perspectives on increased data sharing in health research in the context of the 2023 National Institutes of Health Data Sharing Policy. *PLoS One* 2024; **19**: e0309161.

6   Tan AC, Webster AC, Libesman S, *et al.* Data sharing policies across health research globally: Cross-sectional meta-research study. *Res Synth Methods* 2024; **15**: 1060–71.

7   AI-READI Consortium. AI-READI: rethinking AI data collection, preparation and sharing in diabetes research and beyond. *Nat Metab* 2024; **6**: 2210–2.

8   Owsley C, Matthies DS, McGwin G, *et al.* Cross-sectional design and protocol for Artificial Intelligence Ready and Equitable Atlas for Diabetes Insights (AI-READI). *BMJ Open* 2025; **15**: e097449.

9   Song Z, Ma H, Sun S, Xin Y, Zhang R. Rainbow: reliable personally identifiable information retrieval across multi-cloud. *Cybersecur (Singap)* 2023; **6**: 19.

10  Differences Between PII, Sensitive PII, and PHI. Municipal Websites Central Help Center. https://www.civicengagecentral.civicplus.help/hc/en-us/articles/1500001543581-Differences-Between-PII-Sensitive-PII-and-PHI (accessed July 18, 2024).

11  Department of Health Care Services. List of HIPAA Identifiers. https://www.dhcs.ca.gov/dataandstats/data/Pages/ListofHIPAAIdentifiers.aspx (accessed July 18, 2024).

12  Kayaalp M. Modes of De-identification. *AMIA Annu Symp Proc* 2017; **2017**: 1044–50.

13  Office for Civil Rights (OCR). Health Information Privacy. HHS.gov. 2008; published online May 7. https://www.hhs.gov/hipaa/for-professionals/special-topics/research/index.html (accessed Aug 30, 2024).

14  Office for Human Research Protections (OHRP). Revised Common Rule. HHS.gov. 2017; published online Jan 17. https://www.hhs.gov/ohrp/regulations-and-policy/regulations/finalized-revisions-common-rule/index.html (accessed July 18, 2024).

15  General Data Protection Regulation (GDPR) Compliance Guidelines. GDPR.eu. 2018; published online June 18. https://gdpr.eu/ (accessed July 18, 2024).

16  Chevrier R, Foufi V, Gaudet-Blavignac C, Robert A, Lovis C. Use and Understanding of Anonymization and De-Identification in the Biomedical Literature: Scoping Review. *J Med Internet Res* 2019; **21**: e13484.

17  General Data Protection Regulation (GDPR) – Legal Text. General Data Protection Regulation (GDPR). https://gdpr-info.eu/ (accessed April 20, 2025).

18  Nakayama LF, de Matos JCRG, Stewart IU, *et al.* Retinal Scans and Data Sharing: The Privacy and Scientific Development Equilibrium. *Mayo Clinic Proceedings: Digital Health* 2023; **1**: 67–74.

19  Committee on Strategies for Responsible Sharing of Clinical Trial Data, Board on Health Sciences Policy, Institute of Medicine. Concepts and Methods for De-identifying Clinical Trial Data. In: Sharing Clinical Trial Data: Maximizing Benefits, Minimizing Risk. National Academies Press (US), 2015.

20  Anthem Pays OCR $16 Million in Record HIPAA Settlement Following Largest U.S. Health Data Breach in History. https://www.hhs.gov/guidance/document/anthem-pays-ocr-16-million-record-hipaa-settlement-following-largest-us-health-data-breach (accessed Sept 17, 2024).

21  Wiepert D, Malin BA, Duffy JR, *et al.* Reidentification of Participants in Shared Clinical Data Sets: Experimental Study. *JMIR AI* 2024; **3**: e52054.

22  Wang Z, Kanduri A, Aqajari SAH, *et al.* ECG Unveiled: Analysis of Client Re-identification Risks in Real-World ECG Datasets. 2024; published online Aug 2. http://arxiv.org/abs/2408.10228 (accessed April 20, 2025).

23  Shei R-J, Holder IG, Oumsang AS, Paris BA, Paris HL. Wearable activity trackers-advanced technology or advanced marketing? *Eur J Appl Physiol* 2022; **122**: 1975–90.

24  Canali S, Schiaffonati V, Aliverti A. Challenges and recommendations for wearable devices in digital health: Data quality, interoperability, health equity, fairness. *PLOS Digit Health* 2022; **1**: e0000104.

25  Sancho J, Alesanco Á, García J. Biometric Authentication Using the PPG: A Long-Term Feasibility Study. *Sensors (Basel)* 2018; **18**. DOI:10.3390/s18051525.

26  The Lancet Digital Health. Wearable health data privacy. *Lancet Digit Health* 2023; **5**: e174.

27  Metwally AA, Perelman D, Park H, *et al.* Prediction of metabolic subphenotypes of type 2 diabetes via continuous glucose monitoring and machine learning. *Nat Biomed Eng* 2024; published online Dec 23. DOI:10.1038/s41551-024-01311-6.

28  Britton KE, Britton-Colonnese JD. Privacy and Security Issues Surrounding the Protection of Data Generated by Continuous Glucose Monitors. *J Diabetes Sci Technol* 2017; **11**: 216–9.

29  Martens T, Beck RW, Bailey R, *et al.* Effect of continuous glucose monitoring on glycemic control in patients with type 2 diabetes treated with basal insulin: A randomized clinical trial: A randomized clinical trial. *JAMA* 2021; **325**: 2262–72.

30  Herrero P, Reddy M, Georgiou P, Oliver NS. Identifying Continuous Glucose Monitoring Data Using Machine Learning. *Diabetes Technol Ther* 2022; **24**: 403–8.

31  Kleidermacher D, Klonoff D, Nguyen K, Schwartz N, Xu N. Addressing the Need for Protecting Cybersecurity in Connected Diabetes Devices. IEEE Standards Association. https://standards.ieee.org/beyond-standards/addressing-the-need-for-protecting-cybersecurity-in-connected-diabetes-devices/ (accessed July 18, 2024).

32  Medical Biometric Databases. BioSec.Lab. https://www.comm.utoronto.ca/~biometrics/databases.html (accessed July 18, 2024).

33  Biel L, Pettersson O, Philipson L, Wide P. ECG analysis: a new approach in human identification. *IEEE Trans Instrum Meas* 2001; **50**: 808–12.

34  Chun SY, Kang J-H, Kim H, Lee C, Oakley I, Kim S-P. ECG based user authentication for wearable devices using short time Fourier transform. In: 2016 39th International Conference on Telecommunications and Signal Processing (TSP). IEEE, 2016: 656–9.

35  Pereira TMC, Conceição RC, Sencadas V, Sebastião R. Biometric Recognition: A Systematic Review on Electrocardiogram Data Acquisition Methods. *Sensors* 2023; **23**. DOI:10.3390/s23031507.

36  Wu S-C, Chen P-T, Hsieh J-H. Spatiotemporal features of electrocardiogram for biometric recognition. *Multidimens Syst Signal Process* 2019; **30**: 989–1007.

37  Kim H, Kim H, Chun SY, *et al.* A Wearable Wrist Band-Type System for Multimodal Biometrics Integrated with Multispectral Skin Photomatrix and Electrocardiogram Sensors. *Sensors* 2018; **18**. DOI:10.3390/s18082738.

38  Gim N, Wu Y, Blazes M, Lee CS, Wang RK, Lee AY. A Clinician's Guide to Sharing Data for AI in Ophthalmology. *Invest Ophthalmol Vis Sci* 2024; **65**: 21.

39  Balancing Benefits and Risks: The Case for Retinal Images to Be Considered as Nonprotected Health Information for Research Purposes - 2024. American Academy of Ophthalmology. 2024; published online Jan 1. https://www.aao.org/education/clinical-statement/balancing-benefits-risks-case-retinal-images-to-be (accessed Jan 6, 2025).

40  Szymkowski M, Saeed E, Omieljanowicz M, Omieljanowicz A, Saeed K, Mariak Z. A Novelty Approach to Retina Diagnosing Using Biometric Techniques With SVM and Clustering Algorithms. https://ieeexplore.ieee.org/abstract/document/9134747 (accessed March 9, 2025).

41  Price WN 2nd, Cohen IG. Privacy in the age of medical big data. *Nat Med* 2019; **25**: 37–43.

42  Luna R, Rhine E, Myhra M, Sullivan R, Kruse CS. Cyber threats to health information systems: A systematic review. *Technol Health Care* 2016; **24**: 1–9.

43  Vilhuber L. Reproducibility and transparency versus privacy and confidentiality: Reflections from a data editor. *J Econom* 2023; **235**: 2285–94.

44  Thakkar V, Gordon K. Privacy and Policy Implications for Big Data and Health Information Technology for Patients: A Historical and Legal Analysis. *Studies in health technology and informatics* 2019; **257**. https://pubmed.ncbi.nlm.nih.gov/30741232/ (accessed March 9, 2025).

45  Bai S, Zheng J, Wu W, Gao D, Gu X. Research on healthcare data sharing in the context of digital platforms considering the risks of data breaches. *Front Public Health* 2024; **12**: 1438579.

46  Lin D, McAuliffe M, Pruitt KD, *et al.* Biomedical Data Repository Concepts and Management Principles. *Sci Data* 2024; **11**: 622.

47  Bandrowski A, Grethe JS, Pilko A, *et al.* SPARC Data Structure: Rationale and Design of a FAIR Standard for Biomedical Research Data. bioRxiv. 2021; : 2021.02.10.430563.

48  Markiewicz CJ, Gorgolewski KJ, Feingold F, *et al.* The OpenNeuro resource for sharing of neuroscience data. *Elife* 2021; **10**. DOI:10.7554/eLife.71774.

49  Alper P, Děd V, Herzinger S, *et al.* DS-PACK: Tool assembly for the end-to-end support of controlled access human data sharing. *Sci Data* 2024; **11**: 501.

50  Tryka KA, Hao L, Sturcke A, *et al.* NCBI's Database of Genotypes and phenotypes: dbGaP. *Nucleic Acids Res* 2014; **42**: D975–9.

51  Sudlow C, Gallacher J, Allen N, *et al.* UK biobank: an open access resource for identifying the causes of a wide range of complex diseases of middle and old age. *PLoS Med* 2015; **12**: e1001779.

52  Sydes MR, Johnson AL, Meredith SK, Rauchenberger M, South A, Parmar MKB. Sharing data from clinical trials: the rationale for a controlled access approach. *Trials* 2015; **16**: 104.

53  All of Us Research Program Investigators, Denny JC, Rutter JL, *et al.* The 'All of Us' Research Program. *N Engl J Med* 2019; **381**: 668–76.

54  AI-READI Consortium. Flagship dataset of type 2 diabetes from the AI-READI project. 2024. DOI:10.60775/FAIRHUB.2.

55  Contreras J, Evans B, Hurst S, *et al.* License terms for reusing the AI-READI dataset. DOI:10.5281/zenodo.10642459.

56  Spadaccini A, Beritelli F. Performance evaluation of heart sounds biometric systems on an open dataset. https://ieeexplore.ieee.org/document/6622835 (accessed March 9, 2025).

57  Labati RD, Piuri V, Rundo F, Scotti F, Spampinato C. Biometric Recognition of PPG Cardiac Signals Using Transformed Spectrogram Images. *Pattern Recognition ICPR International Workshops and Challenges* 2021; : 244–57.

58  Retsinas G, Filntisis PP, Efthymiou N, Theodosis E, Zlatintsi A, Maragos P. Person Identification Using Deep Convolutional Neural Networks on Short-Term Signals from Wearable Sensors. https://ieeexplore.ieee.org/document/9053910 (accessed March 9, 2025).

59  Lee E, Ho A, Wang Y-T, Huang C-H, Lee C-Y. Cross-Domain Adaptation for Biometric Identification Using Photoplethysmogram. https://ieeexplore.ieee.org/document/9053604 (accessed March 9, 2025).

60  Hwang DY, Taha B, Da Saem L, Hatzinakos D. Evaluation of the Time Stability and Uniqueness in PPG-Based Biometric System. https://ieeexplore.ieee.org/document/9130730 (accessed March 9, 2025).

61  Yadav U, Abbas SN, Hatzinakos D. Evaluation of PPG Biometrics for Authentication in different states. 2017; published online Dec 22. http://arxiv.org/abs/1712.08583 (accessed March 9, 2025).

62  Tan R, Perkowski M. Toward improving electrocardiogram (ECG) biometric verification using mobile sensors: A two-stage classifier approach. *Sensors* 2017; **17**. DOI:10.3390/s17020410.

63 Arnau-Gonzalez P, Katsigiannis S, Ramzan N, Tolson D, Arevalillo-Herrez M. ES1D: A deep network for EEG-based subject identification. In: 2017 IEEE 17th International Conference on Bioinformatics and Bioengineering (BIBE). IEEE, 2017: 81–5.

64 Zhao Z, Zhang Y, Deng Y, Zhang X. ECG authentication system design incorporating a convolutional neural network and generalized S-Transformation. *Comput Biol Med* 2018; **102**: 168–79.

65 Patro KK, Jaya Prakash A, Jayamanmadha Rao M, Rajesh Kumar P. An efficient optimized feature selection with machine learning approach for ECG biometric recognition. *IETE J Res* 2022; **68**: 2743–54.

66 Patro KK, Reddi SPR, Khalelulla SKE, Rajesh Kumar P, Shankar K. ECG data optimization for biometric human recognition using statistical distributed machine learning algorithm. *J Supercomput* 2020; **76**: 858–75.

67 El Boujnouni I, Zili H, Tali A, Tali T, Laaziz Y. A wavelet-based capsule neural network for ECG biometric identification. *Biomed Signal Process Control* 2022; **76**: 103692.

68 Allam JP, Patro KK, Hammad M, Tadeusiewicz R, Pławiak P. BAED: A secured biometric authentication system using ECG signal based on deep learning techniques. *Biocybern Biomed Eng* 2022; **42**: 1081–93.

69 Prakash AJ, Patro KK, Samantray S, Pławiak P, Hammad M. A deep learning technique for biometric authentication using ECG beat template matching. *Information (Basel)* 2023; **14**: 65.

70 Wang X, Cai W, Wang M. A novel approach for biometric recognition based on ECG feature vectors. *Biomed Signal Process Control* 2023; **86**: 104922.

71 Farzin H, Abrishami-Moghaddam H, Moin M-S. A Novel Retinal Identification System. *EURASIP J Adv Signal Process* 2008; **2008**: 280635.

72 Köse C, İki˙baş C. A personal identification system using retinal vasculature in retinal fundus images. *Expert Syst Appl* 2011; **38**: 13670–81.

73 Biometric Retina Identification Based on Neural Network. *Procedia Computer Science* 2016; **102**: 26–33.

74 Retina biometrics for personal authentication. In: Machine Learning for Biometrics. Academic Press, 2022: 87–104.

75 Marappan J, Murugesan K, Elangeeran M, Subramanian U. Human retinal biometric recognition system based on multiple feature extraction. 2023; published online Jan 20. DOI:10.1117/1.JEI.32.1.013008.

**Table 1**. Glossary of major terms, and concepts relevant in this work.

|  | **Definition** |
|---|---|
| Pseudonymization | The process of replacing private identifiers with fake identifiers, or pseudonyms to protect an individual's identity while retaining data utility. It allows data to be reidentified if necessary using a reidentification key. |
| Anonymization | The process of irreversibly deidentifying data elements, following GDPR** guidelines. |
| Quasi Identifiers | Variables in a research dataset that can not individually identify a participant but, in combination with other variables, identify a record, or participant. |
| Reidentification | The process of matching anonymized, or pseudonymized data with other information, such as a reidentification key, patient ID, publicly available information, and/or other datasets to reestablish the identity of an individual. |
| Deidentification | The process of removing, or altering personal identifiers from data so that the individuals to whom the data pertains cannot be readily identified. Deidentification is often used to maintain privacy in datasets used for research, and analysis. |
| Pseudo-reidentification | The process by which AI, or analytical methods detect unique patterns in deidentified data that suggest individuality without directly linking to a specific person, unlike traditional reidentification, which requires external identifiers, or reference datasets. |
| PII (Personally Identifiable Information) | Any data that could potentially identify a specific individual, including but not limited to names, social security numbers, addresses, phone numbers, and email addresses. |
| PHI (Protected Health Information) | Any health-related information that can be linked to an individual, and is protected under regulations such as HIPAA. PHI includes medical records, insurance information, and other personal health data. |
| HIPAA Covered Entity | Any person, or organization that is authorized to collect, use, and transmit PHI in accordance with HIPAA* regulations. |

\* HIPAA: Health Insurance Portability, and Accountability Act. ** GDPR: General Data Protection Regulation

**Table 2**. List of the 18 HIPAA Safe Harbor Identifiers.

| **HIPAA Identifiers** |
|---|
| Names |

| | |
|---|---|
| All geographic subdivisions smaller than a State, including street address, city, county, precinct, zip code, and their equivalent geocodes, except for the initial three digits of a zip code if, according to the current publicly available data from the Bureau of the Census: <br> 1. The geographic unit formed by combining all zip codes with the same three initial digits contains more than 20,000 people; and <br> 2. The initial three digits of a zip code for all such geographic units containing 20,000, or fewer people is changed to 000 | |
| All elements of dates (except year) for dates directly related to an individual, including birth date, admission date, discharge date, date of death; and all age over 89, and all elements of dates (including year) indicative of such age, except that such ages, and elements may be aggregated into a single category of age 90, or older | |
| Telephone numbers | |
| Fax numbers | |
| Electronic mail addresses | |
| Social security numbers | |
| Medical record numbers | |
| Health plan beneficiary numbers | |
| Account numbers | |
| Certificate/license numbers | |
| Vehicle identifiers, and serial numbers, including license plate numbers | |
| Device identifiers, and serial numbers | |
| Web Universal Resource Locators (URLs) | |
| Internet Protocol address numbers | |
| Biometric identifiers, including finger, and voice prints | |
| Full-face photographic images, and any comparable images | |
| Any other unique identifying number, characteristic, or code | |

**Table 3**. A highlight of the recent publications on the approaches to pseudo-reidentification using health data.

| | Year | Database | Method | Accuracy (%) |
|---|---|---|---|---|
| **Wearable fitness-tracking devices** | | | | |
| Spadaccini, et al.[56] | 2013 | HSCT-11 | GMM | 86·4 |

| Sancho, et al.[25] | 2018 | MIMIC II<br>PRRB | L2 distance | 78·5 |
|---|---|---|---|---|
| Labati, et al.[57] | 2020 | PRRB | SVM | 94·8 |
| Retsinas, et al.[58] | 2020 | PersonID | CNN | 55·8 |
| Lee, et al.[59] | 2020 | IEEEPPG | CNN | 95·7 |
| Hwang, et al.[60] | 2021 | BioSec | CNN + RNN | 87·1 |
| Yadav, et al.[61] | 2021 | BioSec<br>DEAP | LDA | 97·4 |
| **Continuous glucose monitoring devices** | | | | |
| Herrero, et al.[30] | 2021 | REPLACE-BG | SVM | 86·8 |
| **Electrocardiogram** | | | | |
| Tan et al.[62] | 2017 | MIT-BIH<br>PhysioNet<br>Mobile ECG | A two-stage classifier integrating random forest and wavelet distance measure with a probabilistic threshold | 99·52 |
| Arnau-González et al.[63] | 2017 | DREAMER | CNN | 94 |
| Zhao et al.[64] | 2018 | ECG-ID<br>PhysioNet | Generalized S-transformation with CNN | 99 |
| Patro et al.[65] | 2019 | MIT-BIH<br>ECG-ID | Feature Extraction, LASSO, KNN | 99·1 |
| Patro et al.[66] | 2020 | PhysioNet<br>ECG-ID<br>PTBDB | Optimized Feature Selection (GA, PSO, LASSO, EN) with RF | 94·9-95·3 |
| El Boujnouni et al.[67] | 2021 | PTB<br>MIT-BIH | A combination of CWT, DWT, along with a capsule network | 98·1 - 100 |
| Parkash et al.[68] | 2022 | ECG-ID<br>PTB<br>CYBHi<br>UofTDB | A deep learning algorithm based on CNN, and LSTM | 98·2 |
| Parkash et al.[69] | 2023 | ECG-ID | Deep learning | 99·9 |
| Wang et al.[70] | 2023 | ECG-ID<br>MIT-BIH<br>USSTDB | ECG Feature Vector with Pooling Layer for Variable-Length Signals | 91-97·6 |

| | | | **Retinal Images** | | |
|---|---|---|---|---|---|
| Farzin et al.[71] | 2008 | DRIVE STARE | Blood vessel segmentation, Feature generation, Feature matching | 99 | |
| Köse et al.[72] | 2011 | Local data STARE | Vessel segmentation | 95 | |
| Sadikoglu et al.[73] | 2016 | DRIVE | CNN | 97·5 | |
| Szymkowski et al.[40] | 2020 | Local data DRIVE STARE Kaggle Retinopathy | KNN SVM CNN | 96·5 | |
| Devi et al.[73,74] | 2022 | VARIA | ANFIS | 97·2 | |
| Marappan et al.[75] | 2023 | RIDB VARIA DRIVE STARE | Multiple feature extraction | 98 | |

MIMIC-II: Multiparameter Intelligent Monitoring in Intensive Care II, PRRB: Photoplethysmography Respiratory Rate Benchmark, SVM: Support Vector Machine, CNN: Convolutional Neural Network, IEEEPPG: Institute of Electrical, and Electronic Engineers Photopletysmographic Signals Dataset, RNN: Recurrent Neural Network, DEAP: Database for Emotion Analysis using Physiological Signals, LDA: Linear Discriminant Analysis, HSCT-11: Heart Sounds Catania 2011, GMM: Gaussian Mixture Models, MIT-BIH: MIT - Beath Israel Hospital dataset, ECG-ID: ECG Identification dataset, KNN: *K*-nearest neighbour, LASSO: least absolute shrinkage, and selection operator, NSRDB: normal sinus rhythm database, STDB: ST change database, PTB: Physikalisch-Technische Bunde- sanstalt, CYBHi: check your bio-signals here initiative,  UofTDB: the University of Toronto Database, LSTM: long short term memory, CWT: Continuous Wavelet Transform, DWT:  Discrete Wavelet Transform, RF: random forest, EN: elastic net, GA: genetic algorithm, PSO: particle swarm optimization, ANFIS: Adaptive network-based fuzzy inference system.

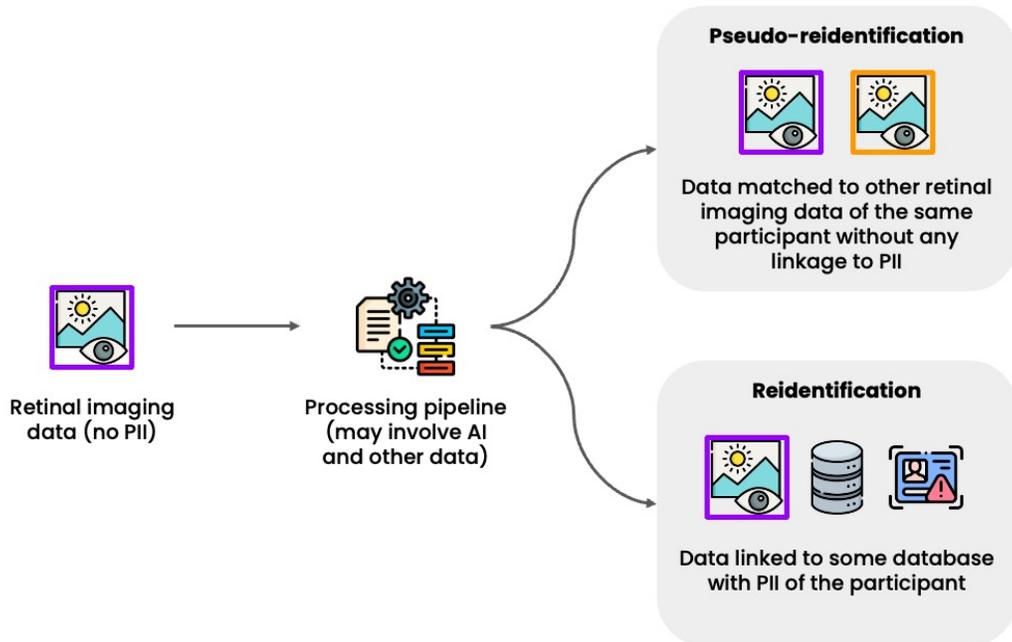

**Figure 1**. Illustration of pseudo-reidentification vs reidentification. AI: artificial intelligence, PII: personal identifier information.

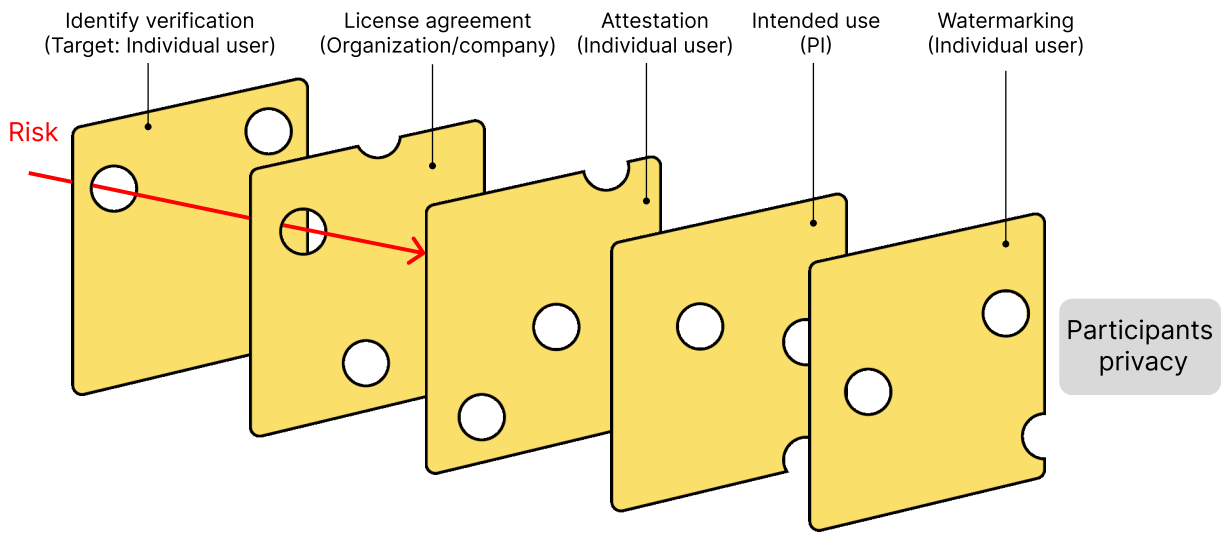

**Figure 2**. Illustration of our new Swiss-cheese method of open data sharing. Each layer is designed to protect participants' privacy from potential risks by targeting primarily the individual user accessing the data, their principal investigator (PI), or their organization/company.